\documentclass{optica-article}

\journal{opticajournal} 

\articletype{Research Article}

\usepackage{lineno}

\usepackage{graphicx}
\usepackage{amsmath}
\usepackage{epstopdf}
\usepackage{color}
\usepackage{lipsum}
\usepackage[colorlinks=true]{hyperref}
\usepackage{ulem}
\usepackage{cleveref}
\usepackage{array}
\usepackage{notes2bib}
\DeclareGraphicsExtensions{.eps}

\newcommand{\bra}[1] {\langle #1|}
\newcommand{\ket}[1] {|#1 \rangle}
\newcommand{\braket}[2] {\langle #1|#2 \rangle}

\newcommand{\opdag}[1]{\hat{#1}^{\dagger}}

\newcommand*{\rom}[1]{\expandafter\@slowromancap\romannumeral #1@}

\usepackage{braket}
\usepackage{graphicx}
\usepackage{dcolumn}
\usepackage{bm}
\usepackage{xcolor}

\begin{document}

\title{Quantum interferences and gates with emitter-based coherent photon sources}

\author{I. Maillette de Buy Wenniger\authormark{1,2,\dag,*}, S. C. Wein,\authormark{3,\dag},\\ D. Fioretto\authormark{2}, S. E. Thomas\authormark{2}, C. Ant\'on-Solanas\authormark{4}, \\ A. Lema\^itre\authormark{1}, I. Sagnes\authormark{1}, A. Harouri\authormark{1}, N. Belabas\authormark{1}, \\N. Somaschi\authormark{3}, P. Hilaire\authormark{3}, J. Senellart\authormark{3} and P. Senellart\authormark{1,3}}

\address{\authormark{1}Universit\'e Paris-Saclay, CNRS, Centre for Nanosciences and Nanotechnology, 91120 Palaiseau, France.\\
\authormark{2}Department of Physics, Imperial College London, London SW7 2BW, UK.\\
\authormark{3}Quandela SAS, 10 Boulevard Thomas Gobert, 91120 Palaiseau, France.\\
\authormark{4}Depto. de F\'isica de Materiales, Instituto Nicol\'as Cabrera, Instituto de F\'isica de la
Materia Condensada, Universidad Aut\'onoma de Madrid, 28049 Madrid, Spain.\\
\authormark{\dag} These authors contributed equally to this work.}

\email{\authormark{*}imaillet@ic.ac.uk} 


\begin{abstract*} 
{Quantum emitters such as quantum dots, defects in diamond or in silicon have emerged as efficient single photon sources that are progressively exploited in quantum technologies. In 2019, it was shown that the emitted single photon states often include coherence with the vacuum component. Here we investigate how such photon-number coherence alters quantum interference experiments that are routinely implemented both for characterising or exploiting the generated photons. We show that it strongly modifies intensity correlation measurements in a Hong-Ou-Mandel experiment and leads to errors in indistinguishability estimations. It also results in additional entanglement when performing partial measurements. We illustrate the impact on quantum protocols by evidencing  modifications in heralding efficiency and  fidelity of two-qubit gates. }
\end{abstract*}

\vspace{1cm}
Single quantum emitters such as atoms, defects in diamond, molecules or quantum dots are natural deterministic single-photon emitters~\cite{Kuhn2002, McKeever2004, Sipahigil2014, Kiraz2005, barros2009, Somaschi2016, Tomm2021}. They can be brought to their excited state and emit single photons with near-unity probability. Their spontaneous emission can be efficiently funnelled into a single optical mode by making use of the Purcell effect when they are inserted in optical cavities.
Together with coherent excitation schemes~\cite{SantoriRF2004, BennettRF2005, WeilerRF2011,Teets1977, Reigue2017, Koong2021, Liu2019}, these methods have allowed the demonstration of single-photon sources with a very high degree of indistinguishability, and efficiencies orders of magnitude higher than heralded single-photon sources based on frequency conversion~\cite{Senellart2017a}, making them great assets to scale-up optical quantum technologies~\cite{OBrien2007}. {In particular quantum dot (QD) based single photon sources are now commercially available and are exploited for quantum computing protocols, enabling a significant increase in the number of manipulated qubits, both on chip and in free space~\cite{Wang2019,Maring2024, Karli2024, Rodari2024, Polacchi2023, Carosini2024, Cao2024}.}

{In 2019 it was shown that, when coherently driven, such quantum emitters can directly generate light wavepackets consisting of arbitrary quantum superpositions of zero and one photon~\cite{Loredo2019}, in other words {\it single-rail qubits}.}  The coherent excitation creates a superposition between the ground and excited state of the two-level system that upon spontaneous emission is transferred to the electromagnetic field. {Such deterministic and efficient generation of single-rail qubits could be of great value for quantum technologies. Indeed, superpositions of zero and one-photon Fock states are a widely studied resource in photonic information processing~\cite{Berry2006SingleRail, Drahi2021} as well as quantum communication protocols~\cite{Liu2013MDIQKD, Rubenok2013QKD, Lo2012MDIQKD}. It is important to note that, so far most of these protocols have been implemented either using weak coherent states~\cite{Liu2013MDIQKD, Rubenok2013QKD, Lo2012MDIQKD} or by interfering single photons with squeezed coherent states~\cite{Berry2006SingleRail, Pegg1998, Babichev2004}. However, such approaches carry strong limitations arising from higher order Fock state components, limiting the scalability of the envisioned protocols~\cite{Berry2007EfficiencyLimit}.}

{The demonstration of deterministic generation of coherent superposition of zero and one photon~\cite{Loredo2019} has recently opened the path towards a more scalable exploitation of single-rail qubits as well as new tailored protocols. Indeed, applications in quantum technologies are progressively emerging, with propositions of application to quantum key distribution~\cite{Bozzio2022}, Boson sampling~\cite{Renema2020}, interfacing single- and dual- rail optical qubits~\cite{Drahi2021} as well as the very first experimental studies~\cite{Polacchi2023}.} 

{In the present work, we revisit Hong-Ou-Mandel (HOM) interference to analyse and understand the effect of coherence with vacuum in linear quantum protocols that are used both for characterising light emitted by quantum emitters as well as for manipulating quantum information. Hence, our work is intended both for the community of researchers developing single-photon sources based on quantum emitters and to those  exploiting these sources for quantum information processing. We first show that coherence with vacuum can lead, and has led, to errors in the measurement of photon indistinguishability for emitter-based single-photon sources. We then explain how it leads to specific quantum interference patterns when performing partial measurements on multiple light pulses, and generates additional entanglement. Considering that partial measurements are central to the scalability of quantum computing, we numerically study the impact of coherence with vacuum on  heralded two-qubit gates. Our findings suggest that tailored quantum information processing protocols can be derived to exploit the deterministic generation photon-number superposition with quantum emitters. }

\vspace{0.5cm}
{The present manuscript is  organised as follows. In Section 1, we recall the formalism of HOM interference with true single-photon states and how it can be used to measure the indistinguishability of the single-photon wavepackets. To provide a reference, we experimentally illustrate this well-known situation using a quantum dot-based single-photon source under incoherent excitation. In Section 2, we study the impact of coherence with vacuum on HOM interference and evidence the emergence of multiple phenomena. We first evidence the observed experimental signatures on a coherently driven quantum dot (2.1) and demonstrate how the intensity correlation measurement should be revised (2.2). Section 3 discusses the errors in indistinguishability measurements that can arise in the characterisation of quantum emitter-based single-photon sources from this overlooked coherence with vacuum. In Section 4 we analyse another feature of HOM interference and evidence how coherence with vacuum leads to additional entanglement when performing partial measurements. Finally, in Section 5 we numerically illustrate how such coherence impacts quantum computing protocols by analysing the case of a heralded control-NOT (CNOT) gate, i.e. a dual-rail encoded protocol fed with wavepackets constituting superpositions of zero and one photons. }

\begin{figure}[t]
\centering
\includegraphics[width=0.6\linewidth]{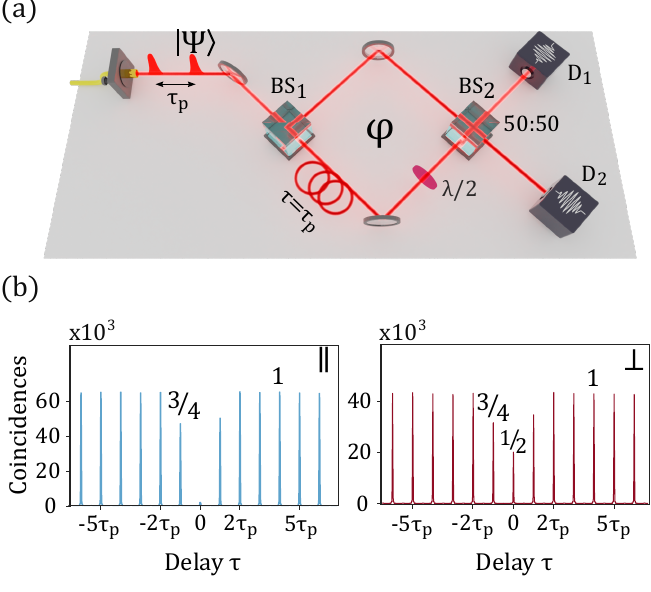}
\caption{(a) The experimental setup to perform Hong-Ou-Mandel interference between two consecutively emitted photonic states ($\tau_p = 12.3$~ns) at a $50:50$ beamsplitter (BS$_2$) where two detectors $\mathrm{D_1}$ and $\mathrm{D_2}$ measure coincidences at the output. The photonic states pick up a relative phase $\varphi$ when passing through the interferometer. The half waveplate $\lambda/2$ determines parallel or orthogonal polarisation of the two arms in the interferometer. (b) The coincidence histograms for parallel ($\parallel$) and perpendicular ($\perp$) polarisation interference configuration for incoherent pulsed excitation of the emitter via phonon-assisted excitation, so that coherence with vacuum is negligible.
\label{fig:1}}
\end{figure}

\section{\label{sec:standardHOM} {Hong-Ou-Mandel interference for indistinguishability measurements}}

{Quantifying the degree of indistinguishability of photons emitted by an (artificial) atom is critical to develop useful quantum light sources for quantum technologies. A typical test is the ability of the quantum emitter to successively emit identical single-photon pulses. We illustrate the typical implementation for performing such indistinguishability measurements in Fig.~\ref{fig:1}(a)}. A stream of light pulses separated in time by $\tau_p$, with $1/\tau_p$ being the repetition rate of the excitation laser, are sent to the input of a path-unbalanced Mach-Zehnder interferometer, where delayed wavepackets are temporally overlapped at the final beamsplitter BS$_2$. Single-photon detectors, D$_{1}$ and D$_{2}$ monitoring the two outputs, register both single and coincidence counts. The relative polarisation of the photons entering the last beamsplitter is controlled by a half waveplate ($\lambda/2$) on one input of BS$_2$. Two measurements are then performed: one with parallel ($\parallel$) polarisation for both input fields, i.e. where the photons are made as identical as possible in all degrees of freedom, and one with orthogonal ($\perp$) polarisation where no interference takes place. 

{We recall} the standard experimental features of HOM interference with single-photon states. To do so, we use a semiconductor quantum dot (QD) inserted in a microcavity pillar as described in Refs.~\cite{Giesz2016, Somaschi2016}. { We use here an excitation scheme that does not generate coherence with vacuum. It relies on phonon-assisted excitation that has been shown to maintain near-unity indistinguishability while providing an incoherent population transfer of the quantum emitter~\cite{BarthLA2016, CosacchiLA2019, Thomas2021}}

Fig.~\ref{fig:1}(b) presents the two detector cross-correlation histograms in parallel and orthogonal configuration under such excitation. Each peak in these histograms corresponds to a correlation measurement with respect to different delays between photons detected by the two detectors, $\tau=k\tau_p$ for integer $k$. The absence of coincidence counts in the $k=0$ delay peak for the parallel configuration indicates the quantum interference of perfectly indistinguishable single photons, where both photons bunch at the output of the final beamsplitter. When the indistinguishability is imperfect, the area of the zero delay peak gives access to the mean wavepacket overlap $M$, quantifying the indistinguishability. 

In the limit of high loss, the coincidence histograms give access to intensity correlation functions, since the probability of detection becomes proportional to the average photon-number at the detector. The area of the zero delay peak in the parallel polarisation configuration is proportional to 
\begin{equation}
    G^{(2)}_{\mathrm{D_1, D_2}, \parallel} (k=0)= \frac{1}{4} \iint 
    G^{(2)}_\mathrm{D_1,D_2, \parallel}(t_1, t_2)dt_1 dt_2,
\end{equation}
where $G^{(2)}_{\mathrm{D_1,D_2}, \parallel}(t_1, t_2)=\braket{\opdag{a}_1(t_1)\opdag{a}_2(t_2)\hat{a}_2(t_2)\hat{a}_1(t_1)}$ is the unnormalised two-time second-order intensity correlation function for detectors $\mathrm{D_1}$ and $\mathrm{D_2}$ with detection times $t_1$ and $t_2$ monitoring the output modes described by photon annihilation operators $\hat{a}_1$ and $\hat{a}_2$, respectively. The integrals are taken over the duration of a single pulse. The \textit{normalised} correlation function then reads:
\begin{equation}
    g_{\mathrm{D_1,D_2}, \parallel}^{(2)}(k=0)=\frac{4}{\mu^2} G^{(2)}_{\mathrm{D_1, D_2},  \parallel} (k=0), \label{eq_area_zero_peak}
\end{equation}
where $\mu/2$ is the average photon-number in each input port of the final 50:50 beamsplitter BS$_2$. 

In practice, the experimental normalisation procedure provides absolute values of intensity correlation functions without knowledge on the transmission and efficiency of every component in the experimental setup - a difficult task that falls into the category of metrology. One way to obtain the normalised intensity correlation function is to integrate the coincidences over the duration of the measurement and subsequently normalising by the product of the total single detection probabilities measured by each detector. Equivalently, one can simply normalise the area of the zero delay peak by the area of the far delay peaks of the correlation histogram either in parallel or perpendicular configuration $G^{(2)}_{\mathrm{D_1, D_2}, \parallel(\perp)}(|k|\geq 2)$ since: 
\begin{align}
G^{(2)}_{\mathrm{D_1, D_2}, \perp}(|k|\geq 2)&= G^{(2)}_{\mathrm{D_1, D_2}, \parallel}(|k|\geq 2) \nonumber \\
&= \iint I_1(t_1)I_2(t_2)dt_1dt_2 =\frac{\mu^2}{4}, \label{eq:nocoh}
\end{align}
where $I_i(t)=\braket{\opdag{a}_i(t)\hat{a}_i(t)}$ is the average intensity at detector $\mathrm{D_i}$, and $\mu_i=\mu/2$ for balanced beamsplitters BS$_1$ and BS$_2$. After normalisation, the area of the coincidence peaks for $|k|\geq 2$ is 1 and the single-photon indistinguishability $M$ is obtained via the HOM visibility defined as~\cite{trivedi2020, Ollivier2021}
\begin{equation}
V_\mathrm{HOM} = 1 - \frac{g^{(2)}_{\mathrm{D_1, D_2}, \parallel}(k=0)}{g^{(2)}_{\mathrm{D_1, D_2}, \perp}(k=0)},
\label{eq:g2nocoh}
\end{equation}
where $g_{\mathrm{D_1, D_2}, \parallel(\perp)}^{(2)}(k=0)$ is the \textit{normalised} second-order intensity correlation for zero delay between detectors in parallel (orthogonal) configuration. As an example, from the experimental data shown in Fig.~\ref{fig:1}(b), we deduce a total mean wavepacket overlap $M=V_\mathrm{HOM} =(91.66 \pm 0.26)\%$ using this approach. 

A last typical feature of the normalised experimental peaks is the area of the $|k|=1$ peak that amounts to 3/4. This can be understood considering the number of ways a pulse sequence of single photons can contribute to each peak~\cite{Loredo2016}. There are three unique paths for the photons to take that contribute to the $|k|=1$ peak, each with the same probability of occurring when considering balanced beamsplitters. However, for the $|k|\geq 2$ peaks, there are four possible paths, and hence the observed normalised peak ratio of 3/4.

\section{\label{sec:HOMcoherence} Impact of coherence with vacuum on Hong-Ou-Mandel interference \protect}

{We now revisit HOM interference with single-photon wavepackets showing quantum coherence with vacuum. We first evidence the multiple experimental signatures arising from coherence with vacuum, provide the theoretical framework to account for these observation, and explain how to adapt the experimental protocol accordingly.}

\subsection{\label{sec:signatures}Experimental signatures \protect}

\begin{figure}[t]
\centering
\includegraphics[width=0.6\linewidth]{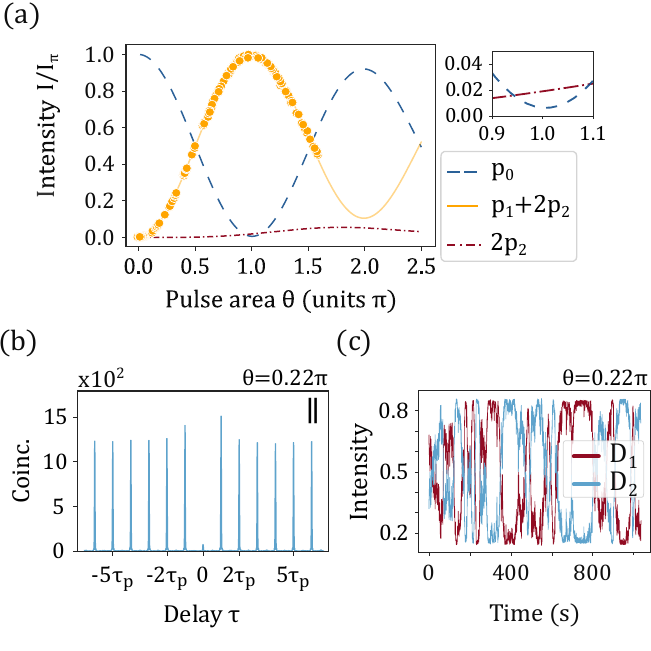}
\caption{(a) The measured QD emission intensity (yellow dots) as a function of pulse area $\theta$ of the resonant driving laser shows the onset of Rabi oscillations. Also shown are the theoretical fit to the data (yellow line) using a simple two-level system model, and the corresponding predictions for zero-photon probability (blue dashed) and the two-photon contribution to the emission intensity (red dashed-dotted). The curves are normalised to the maximum emission intensity $I_\pi$ (at $\theta=\pi$). The inset shows the zero- and two-photon predictions near $\theta=\pi$. (b) The two-time second-order correlation histogram for an emitter driven with pulse area $\theta=0.22\pi$ in parallel polarisation configuration. (c) The intensities measured by detectors $\mathrm{D_1}$ and $\mathrm{D_2}$ in a Hong-Ou-Mandel setup for a resonantly driven emitter with pulse area $\theta=0.22\pi$, showing anticorrelated oscillations as a function of the freely evolving phase $\varphi$.
 \label{fig:2}}
\end{figure}

{ To generate a quantum superposition of zero and one photon, we exploit a resonant excitation scheme,} the most widely used excitation technique to obtain indistinguishable single-photons. In practice, the laser is resonant to the QD transition and the emitted light is separated from the laser using a cross-polarisation configuration~\cite{Somaschi2016}. Fig.~\ref{fig:2}(a) shows the emission intensity $I$ as a function of pulse area $\theta$ (yellow circles) normalised by the emission intensity at $\theta=\pi$, evidencing the onset of Rabi oscillations. 
Fig.~\ref{fig:2}(b) shows the coincidence peaks in parallel configuration measured for pulse area $\theta=0.22\pi$. While the coincidence peaks for $\theta=\pi$ appear close to the case of incoherent excitation (not shown), distinct differences in relative peak heights are observed for $\theta=0.22\pi$. Notably, the area of the $|k|=1$ peaks is greater than the area of the $|k|\geq 2$ peaks. 
Another important signature of photon-number coherence is observed when considering the single counts on each detector \cite{Loredo2019}. Fig.~\ref{fig:2}(c) evidences that the single counts slowly fluctuate over time in opposite phase, indicating the presence of substantial photon-number coherence for $\theta=0.22\pi$.

As demonstrated in Ref.~\cite{Loredo2019}, for sources with negligible multi-photon emission  the coherence between the ground and excited state imprinted by the laser is transferred to the electromagnetic field through spontaneous emission~\bibnote{For simplicity, we describe our system in the following as an effective two-level system with instantaneous state preparation, implying $g^{(2)}(0)=0$.}. The resulting state of light is then well described by:
\begin{equation}
    \ket{\Psi(\theta,\alpha)}=\cos{\left( \frac{\theta}{2}\right)}\ket{0} + e^{i\alpha}\sin{\left( \frac{\theta}{2}\right)}\ket{1},  
    \label{ref-eq1}
\end{equation}
where the phase $\alpha$ is imposed by the laser. The pulse area allows for tuning of the zero- ($p_0=\cos^2(\theta/2)$) and one- ($p_1=\sin^2(\theta/2)$) photon populations and photon-number coherence $\rho_{01}=e^{i\alpha}\sin{\left( \frac{\theta}{2}\right)}\cos{\left( \frac{\theta}{2}\right)}$.  In our experiment, the interferometer is not actively stabilised and the two wavepackets pick up a slowly varying relative phase $\varphi$, i.e. the amplitude of light in one arm experiences a phase shift $\alpha \rightarrow\alpha+\varphi$, leading to phase-dependent quantum interference at beamsplitter BS$_2$. The single counts $I_{1,2}$ measured by detectors D$_{1,2}$ are now proportional to:
\begin{equation}
    I_{1,2}\propto \frac{\mu}{2}\left(1\pm c^{(1)} \cos{(\varphi)}\right),
    \label{ref-eq2}
\end{equation}
where $c^{(1)}=p_0=\cos^2(\theta/2)$ for a state comprising only $0$ or $1$ photon.  More generally, for pure photonic states of the form $\ket{\Psi}=\sum_{n=0}^\infty\sqrt{p_n}\ket{n}$ this quantity is:
\begin{equation}
c^{(1)}=\frac{1}{\mu}\left|\sum_{n=0}^\infty\sqrt{(n+1)p_np_{n+1}}\right|^2,
\label{eq:c1}
\end{equation}
and it quantifies the mean first-order photon-number coherence between states containing $n$ and $n+1$ photons. The oscillations in single counts reflect single-photon interference phenomena that take place between two indistinguishable emitted wavepackets. Importantly, this interference effect invalidates the standard normalisation procedures described in the previous section: those relying on the recorded single counts or on the areas of far delay peaks $|k|\geq2$.

\subsection{\label{sec:norm}Normalisation in the presence of coherence with vacuum \protect}

In the presence of coherence with vacuum, Eq.~\ref{ref-eq2} shows that in parallel configuration, the product $I_1I_2$ depends on $\varphi$, and the areas of the $|k|\geq 2$ delay peaks are now given by  
\begin{equation}
\label{eq:prop2tau}
    G^{(2)}_{\mathrm{D_1, D_2}, \parallel}(|k|\geq 2) \propto 1- \left(c^{(1)} \cos{(\varphi)}\right)^2. 
\end{equation}
This is a manifestation of first-order interference, similar to classical interference, where the interferometer phase causes fringes at the output that can increase the counts at one detector while decreasing the counts at the other. This results in an overall reduced coincidence count rate unless $\varphi=\pi/2$ exactly. As a consequence, when overlooking the presence of this type of quantum interference and having no control over the phase $\varphi$, the far delay peaks will be smaller than expected when $c^{(1)}\neq 0$. Hence, the normalisation factor $\mu^2/4$, that should be independent of phase and coherent effects, can no longer be obtained from just the far delay peaks of the detector cross-correlation histogram \bibnote{This normalisation method is also incorporated as a standard feature in Ref.~\cite{Swabian2023}.}.

As a solution, one can access the normalisation factor by recording both the cross-  and auto-correlation functions $G^{(2)}_{\mathrm{D}_i, \mathrm{D}_j} (t_1, t_2)$ with $i\neq j$ and $i= j$, respectively, considering that:
\begin{equation}
    \frac{\mu^2}{4} = \frac{1}{4}\left( G^{(2)}_{\mathrm{D_1, D_1}} (k) + 2G^{(2)}_{\mathrm{D_1, D_2}} (k) + G^{(2)}_{\mathrm{D_2, D_2}} (k)\right),
    \label{eq:norm}
\end{equation}
with $|k| \geq 2 $ and assuming parallel polarisation. This normalisation factor is also conveniently robust against efficiency imbalances in the interferometer (see Supplemental). \\

\begin{figure*}[t]
\centering
\includegraphics[width=\linewidth]{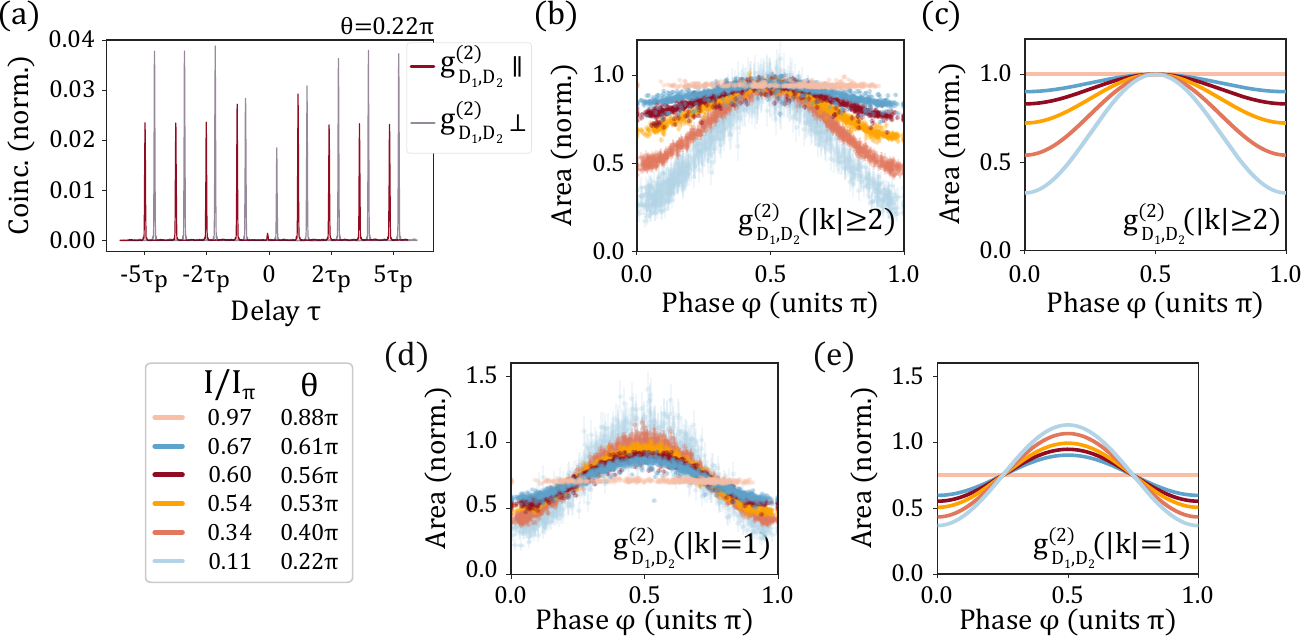}
\caption{ (a) The time-integrated coincidence histograms in parallel (red) and orthogonal (grey) polarisation configuration for $\theta=0.22\pi$ normalised using the phase-independent normalisation factor (see main text). The delay axis of the grey curve is shifted for clarity. (b) The experimentally extracted normalised areas of the $|k|\geq 2$ peaks from phase-resolved coincidence histograms $g_{\mathrm{D_1, D_2}, \parallel}^{(2)}$ as a function of optical phase $\varphi$ and pulse area $\theta = 2 \arcsin{(I/I_\pi)}$. (c) The theoretical prediction of the normalised far delay peak areas ($|k|\geq 2$) as a function of optical phase $\varphi$ and pulse area. (d) and (e) the same as in (b) and (c), respectively, but for the $|k|=1$ peak areas.  \label{fig:3}}
\end{figure*}

Using this phase-independent and coherence-robust normalisation procedure, we plot in Fig.~\ref{fig:3}(a) the normalised coincidence histograms in parallel (red) and orthogonal (grey) polarisation configuration for $\theta=0.22\pi$. We observe a strong suppression of the far delay peaks as anticipated by the proportionality in Eq.~\ref{eq:prop2tau}, here averaged over the fluctuating phase $\varphi$. We also perform phase-resolved intensity correlation measurements using time-tagging techniques, where we divide the data into separate normalised histograms depending on the instantaneous interferometer phase $\varphi$ deduced from the detector single counts (see Supplemental for experimental details). 
Fig.~\ref{fig:3}(b) shows the experimentally measured peak areas of the $|k|\geq 2$ delay peaks as a function of $\varphi$ for various probed pulse areas $\theta$. An increasingly strong phase-dependence is observed as the vacuum component of the photonic state increases. These observations agree well with the theoretical prediction 
\begin{equation}
    g_{\mathrm{D_1,D_2},\parallel}^{(2)}(|k|\geq2)=1- \left(c^{(1)} \cos{(\varphi)}\right)^2,
    \label{eq:2tau}
\end{equation}
as illustrated in Fig.~\ref{fig:3}(c).

Conversely, we show in the Supplemental, experimentally and theoretically, that the normalised zero delay peak $g^{(2)}_{\mathrm{D_1, D_2}, \parallel} (k=0)$ exhibits no phase-dependence. One can thus use the standard relation (Eq.~\ref{eq:g2nocoh}) along with this phase-independent normalisation procedure to extract the indistinguishability, even in the presence of coherence with vacuum.

We also analyse the area of the $|k|=1$ peaks of the properly normalised histograms as shown in Fig.~\ref{fig:3}(d) which evidences strong phase dependence increasing with the vacuum component. When $\theta\simeq\pi$ we observe a normalised area close to $3/4$, but for $\theta<\pi$ the peak areas can exhibit both lower and higher values. Through a complete theoretical analysis of the path-unbalanced interferometer including coherent effects (see Supplemental) we find that the normalised $|k|=1$ peak area is described by 
\begin{equation}
    g_{\mathrm{D_1,D_2},\parallel}^{(2)}(|k|=1)=\frac{1}{4} +\frac{1}{2}\left(1- s^{(2)}_{\{1|M\}}\cos(2\varphi)\right),
    \label{eq:1tau}
\end{equation}
where the amplitude $s^{(2)}_{\{1|M\}}$  quantifies the joint temporal overlap between the first-order photon-number coherence $\braket{\hat{a}(t)}$ that dictates $c^{(1)}$, and the first-order two-time amplitude correlation $\braket{\hat{a}^\dagger(t_1)\hat{a}(t_2)}$ that dictates $M$. Similar to the $g^{(2)}$ notation, the superscript $(2)$ in $s^{(2)}_{\{1|M\}}$ indicates here that this amplitude is a second-order correlation, i.e. a two-photon process. In the case of an emitter subject to pure dephasing only, we theoretically show that:
\begin{equation}
    s^{(2)}_{\{1|M\}} = c^{(1)}\left(\frac{2M}{1+M}\right).
    \label{eq:joints1M}
\end{equation}
 Fig.~\ref{fig:3}(e) illustrates the theoretically expected behaviour for perfect single-photon indistinguishability ($M =1$) as a function of pulse area $\theta$ and interferometer phase $\varphi$ accurately accounting for our experimental observations.

{The above observations and theoretical analysis evidence how the quantum interference of single wavepackets on a beamsplitter is modified in the presence of coherence with vacuum. In the next section, we discuss how these changes should be taken into account for accurate measurements of the indistinguishability of single-photon wavepackets from quantum emitters. }
 
\section{\label{sec:ErrorM} Errors in indistinguishability measurements \protect}

Many quantum emitters are investigated as sources of indistinguishable photons (atoms, ions, semiconductor QDs, defects in 2D materials etc.)~\cite{aharonovich2016}.  
When pursuing the generation of indistinguishable photons, coherent control schemes are naturally adopted as they ensure the lowest degree of time jitter for the spontaneous emission process. 

{ Due to its relatively recent evidence~\cite{Loredo2019}, the presence of photon-number coherence in the emission of resonantly excited atoms or artificial atoms has been widely ignored so far. As a result, the influence of optical phases in the experimental apparatus that play an important role in the presence of coherence with vacuum (see Section 2), has been completely overlooked. Importantly, we underline that  such phase-resolved analysis would add} a great level of complexity to the experimental study, requiring active phase-stabilisation or high photon collection efficiency to trace the phase effect as it varies within the measurement time, as well as event timing to access auto-correlation signals. {Still, we show now that there is actually a simple way to identify the presence of photon-number coherence in standard experimental studies of photon indistinguishability, which allows to trace back errors in its estimation}.\\

\begin{figure}[t!]
\centering
\includegraphics[width=0.6\linewidth]{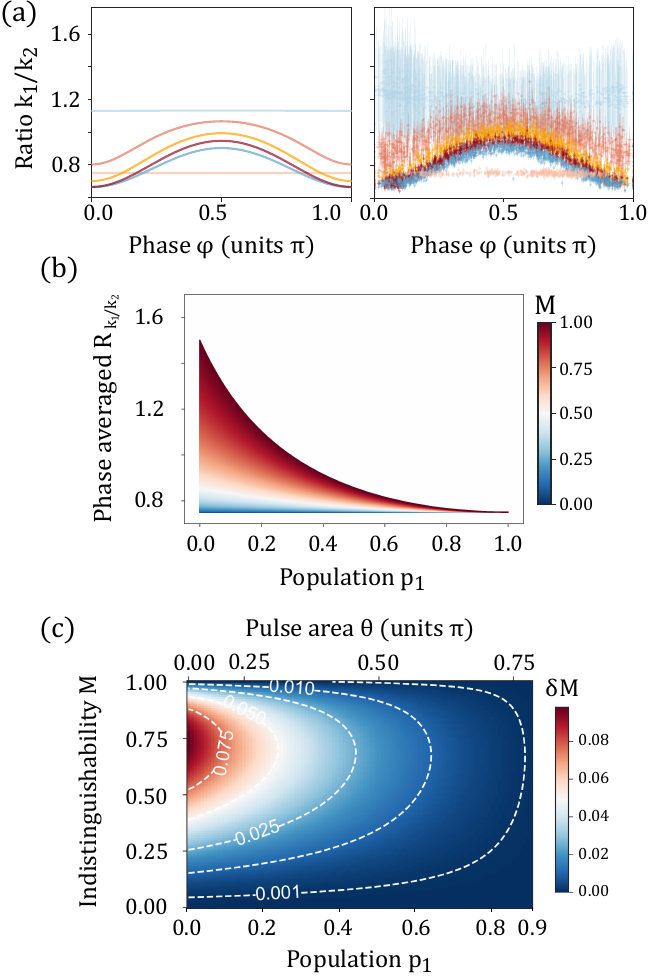}
\caption{(a) The ratio of the normalised $|k|=1$ and $|k|\geq2$ peak areas (left: theoretical, right: experimental) as a function of optical phase $\varphi$ for different pulse areas $\theta$. The colours follow the same colour map as in Fig.~\ref{fig:3}.  (b) The phase averaged peak ratios of the normalised $|k|=1$ and $|k|\geq2$ peak areas as a function of excited state population $p_1$ and indistinguishability $M$ (colour coded) for a purely dephased emitter. (c) The error in indistinguishability $\delta M$ as a function of pulse area $\theta$ (or $p_1$) and single-photon indistinguishability $M$ considering a purely dephased emitter.
\label{fig:4}}
\end{figure}

We consider the ratio $R_{k_1/k_2}$ of the $|k|=1$ and $|k|\geq 2$ peak areas. In Fig.~\ref{fig:4}(a) we plot this ratio as a function of the interferometer phase $\varphi$ for both the theoretical prediction and our experimental data, once again illustrating good agreement between the two. If the measurements are not phase-resolved but phase-averaged, $\cos(2\varphi)$ in Eq.~\ref{eq:1tau} vanishes and $\cos^2(\varphi)$ in Eq.~\ref{eq:2tau} tends to 1/2. Thus the ratio becomes $R_{k_1/k_2} = 3/(4 - 2(c^{(1)})^2)$ and $R_{k_1/k_2}>3/4$ implies that $c^{(1)}>0$, hence indicating the presence of first-order coherence. We also note that this technique to measure $c^{(1)}$ can work even if the source has non-negligible multi-photon emission by applying a minor correction to $R_{k_1/k_2}$.

Fig.~\ref{fig:4}(b) shows the theoretically calculated ratio assuming phase averaging and a purely dephased emitter as a function of $\theta$ for various $M$, evidencing a correspondence between the measured ratio and the fraction of coherent vacuum in the photon state. We can thus give an estimation for the population of coherent vacuum contributing to previous measurements in the literature. As an example, we have gathered in Table~\ref{tableM} the ratio $R_{k_1/k_2}$ estimated from experimental data from some works ~\cite{HOM_dalio2022, HOM_He2017, HOM_Matthiesen2013, HOM_Gerhardt2019, Karli2023, Ourari2023} where a clear deviation from the $3/4$ value is observed. We deduce the corresponding estimation of vacuum population present (column 4 and 5) illustrating how ignoring the presence of coherence with vacuum has led to errors in the derived indistinguishability values. Interestingly, the errors lead to an underestimation of the photon indistinguishability. 

Fig.~\ref{fig:4}(c) shows the calculated error $\delta M$ as a function of pulse area $\theta$ for the phase-averaged scenario. Here, our study is limited to the case where the two-photon component is negligible, typically for values $p_1<0.9$. We define this error in indistinguishability as $\delta M= M - V_\mathrm{HOM}$ where we consider $M$ the single-photon indistinguishability and $V_\mathrm{HOM}$ the indistinguishability extracted from coincidence histograms using the areas of the far delay peaks as a reference. A tentative estimation of the errors in the literature is shown in the last column of Table~\ref{tableM}. These errors appear small, but we underline that every fraction of a percent is critical when optimising the indistinguishability of source emission - a requirement for fault tolerant quantum information processing.\\

\setlength{\tabcolsep}{0.3em}
\begin{table}
\centering
\begin{tabular}{ | c | c | c | c | c | c | }
\hline
 \textbf{Emitter} & \textbf{Ref.} & \textbf{Reported} $\mathbf{V_\mathrm{HOM}}$ &  $\mathbf{R_{k_1/k_2}}$ & $\mathbf{p_0}$  & $\mathbf{\delta M}$ \\ \hline 
QD& \cite{HOM_dalio2022} & $(89.2\pm0.9)\%$ & $0.90\pm0.06$ & $40\%$ &$1\%$ \\  
QD &\cite{HOM_He2017} & $(91.1\pm1.9)\%$ & $0.80\pm0.04$  & $30\%$ & $0.5\%$   \\
QD &\cite{HOM_Matthiesen2013} & $(92.6\pm1.6)\%$ & $0.97\pm0.04$  & $55\%$ &$1.5\%$ \\
QD &\cite{HOM_Gerhardt2019} & $(93.0\pm1.3)\%$ & $1.04\pm0.04$  & $60\%$ & $1.5\%$\\
QD &\cite{Karli2023} & $(95\pm4)\%$ & $0.88 \pm 0.04$  & $30\%$ &$0.5\%$\\
Ion &\cite{Ourari2023} & $(80\pm4)\%$ & $0.86\pm0.02$  & $30\%$& $0.5\%$ \\ \hline
 \end{tabular}
 \caption{Estimated ratio $R_{k_1/k_2}$ in a selection of prior publications resulting in an estimation for the vacuum component $p_0$ and errors on indistinguishability.}\label{tableM}
\end{table}

Note that we have so far focused our study on the coherence between the zero- and one-photon components, which  holds below $\pi$-pulse and for short enough excitation pulses. However, it is important to emphasise that the above-described effects can also occur close to and beyond $\theta=\pi$ where the first-order photon-number coherence may appear between higher photon-number components following Eq.~\ref{eq:c1}. Fig.~\ref{fig:2}(a) shows the theoretical predictions of the vacuum probability $p_0$ and the one $p_1$ and two-photon contribution $2p_2$  along the Rabi curve upon resonant excitation with a finite pulse of $7$~ps duration and an emission decay time of approximately $161$~ps. The two-photon component (dashed-dotted line) is expected to be significantly larger than the zero-photon component at $\theta=\pi$. This effect still results in a non-zero first-order coherence at maximum excited state population (i.e. maximal brightness of the source),  mainly dictated by the coherence between the one- and two-photon component according to Eq.~\ref{eq:c1}. Experimentally, we indeed witness the presence of photon-number coherence in anti-correlated oscillations in the single counts, see Supplemental Fig. S2. Thus, corrections to indistinguishability measurements must also be implemented at the highest brightness of the source i.e. around $\theta=\pi$ by including the analysis of the emitted state up to two photons.

\section{\label{sec:quantumint} Entanglement in partial measurements \protect}

{In this section we discuss the quantum phenomena behind the observation of phase-dependent single counts (see Fig.~\ref{fig:2}(c)). We show that it arises from partial photonic measurements and reveals the presence of spatio-temporal entanglement.}\\

\begin{figure*}[t]
\centering
\includegraphics[width=\linewidth]{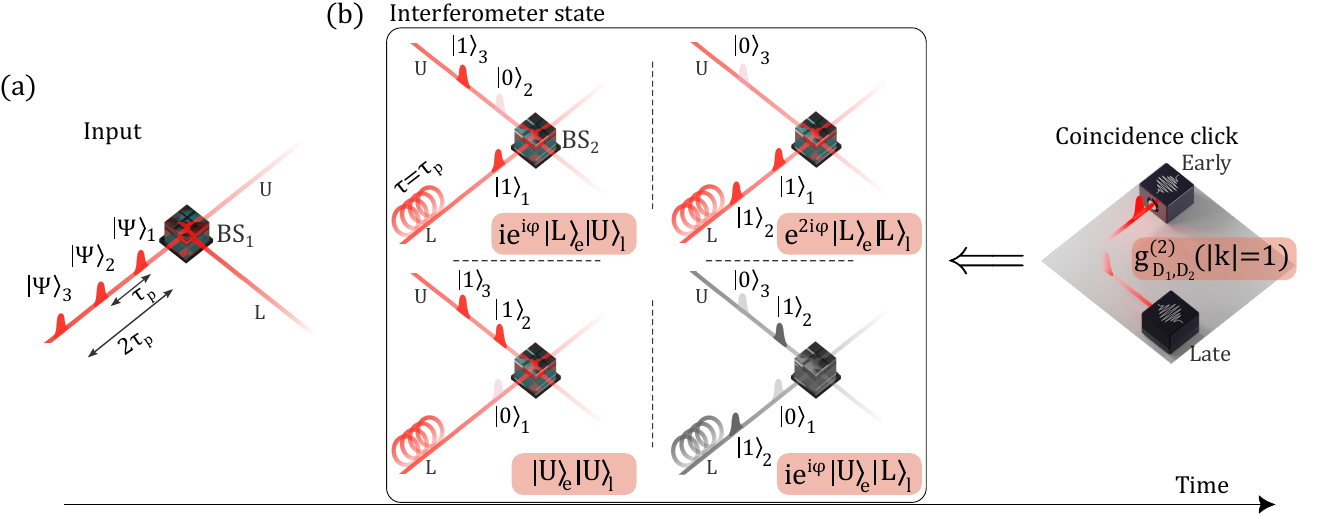}
\caption{ (a) The considered input to the unbalanced Mach-Zehnder interferometer. $U$ and $L$ denote the upper and lower path of the interferometer.  (b) The detection of $g_{\mathrm{D_1,D_2},\parallel}^{(2)}(|k|=1)$ coincidence clicks (separated in time by $\tau=\tau_p$) projects the interferometer state into the path and time entangled state shown on the left (at the interfering beamsplitter BS$_2$). The subscripts $e, l$ denote early and late detection, respectively. The state of light $ie^{i\varphi}\ket{U}_e\ket{L}_l$ inside the interferometer can only be produced by an input state involving two photons in one pulse, and hence it has a negligible chance of occurring. See main text for details.
\label{fig:5}  }
\end{figure*}

We first consider the events leading to a coincidence for $|k|=1$ {in the unbalanced Mach-Zehnder interferometer with a delay line ($\tau = \tau_p$) implemented in one of the arms}. In the high loss regime, a $|k|=1$ coincidence count implies the detection of exactly two photons, one at each detector {and separated in time by $\tau_p$ in an early and late detection bin}. Fig.~\ref{fig:5}(a) shows the pulse sequence arriving at the first beamsplitter of the MZ interferometer that contribute to this signal - each pulse at the input of the interferometer is a quantum superposition of 0 and 1 photon (Eq.~\ref{ref-eq1}). Labelling $\ket{U}_{e,l}$ and $\ket{L}_{e,l}$ a single-photon arriving from the upper (U) and lower (L) input of the last beamsplitter, in early (e) or late (l) detection time bins, there are four states of light impinging on BS$_2$ that lead to a two-photon coincidence count contributing to $g^{(2)}_{\mathrm{D_1,D_2}, \parallel}(|k|=1)$: $\ket{U}_e\ket{U}_l$, $\ket{U}_e\ket{L}_l$, $\ket{L}_e\ket{U}_l$, and $\ket{L}_e\ket{L}_l$, see Fig.~\ref{fig:5}(b). The case $\ket{U}_e\ket{L}_l$, however, can only result from light pulses containing more than one photon - a situation that has a negligible chance of occurring in our experiment and so we disregard it.

If the remaining three cases are indistinguishable, the detection of $|k|=1$ coincidences actually projects the state of light entering BS$_2$ onto the partially path-entangled state:
\begin{equation}
    \!\!\!\!\ket{\Psi}=\frac{1}{\sqrt{3}}\!\left(ie^{i\varphi}\ket{L}_e \!\ket{U}_l+\ket{U}_e\!\ket{U}_l-e^{2i\varphi}\ket{L}_e\!\ket{L}_l\right)\!,\!
    \label{eq:path_time_entangle}
\end{equation}
a state with an entanglement concurrence of $C=2/3$~\cite{Wootters2001}. In practice, photon losses and photon distinguishability will cause these states to become partially distinguishable, which reduces the amount of path entanglement and hence reduces the visibility of oscillations in the $|k|=1$ peak areas. For example, a three-photon state followed by the loss of a single photon could produce the outcome $\ket{L}_e\ket{U}_l$ while being completely distinguishable from the other two cases via the lost photon (see Supplemental). 

We find that the amplitude $s^{(2)}_{\{1|M\}}$, quantifying the visibility of the $|k|=1$ coincidences, is related to the entanglement concurrence of this path-entangled state entering BS$_2$, according to $C=(2/3)s^{(2)}_{\{1|M\}}$.  The observation of phase-dependent areas of the $|k|=1$ peaks thus reveals the generation of spatio-temporal entanglement with a maximal concurrence of $2/3$ when $\theta$ tends to zero and $M$ to one, that conversely vanishes when approaching $\theta=\pi$.

\section{\label{sec:heraldedGates} Impact on heralded gates \protect}

{Based on the above study of HOM interference, we are now in position to reach} some general understanding about the impact of coherence with vacuum on linear quantum protocols. Starting from $N$ pulses containing quantum superpositions of zero and one photons, we expect no modification of linear quantum processing protocols when detecting $N$ single-photons. In such case the measurement post-selects on all pulses being in the Fock state 1. However, large scale linear quantum computing relies on partial measurement of photons and feed-forward. In such case, by measuring only $n$ photons out of $N$ pulses, one does not post-select on a single configuration, but on multiple interfering quantum trajectories where the $n$ photons come from different pulses. Such additional quantum interferences will modify the operation of any heralded protocol differently from the presence of photon loss (incoherent vacuum).\\

\begin{figure*}[t!]
\includegraphics[width=\linewidth]{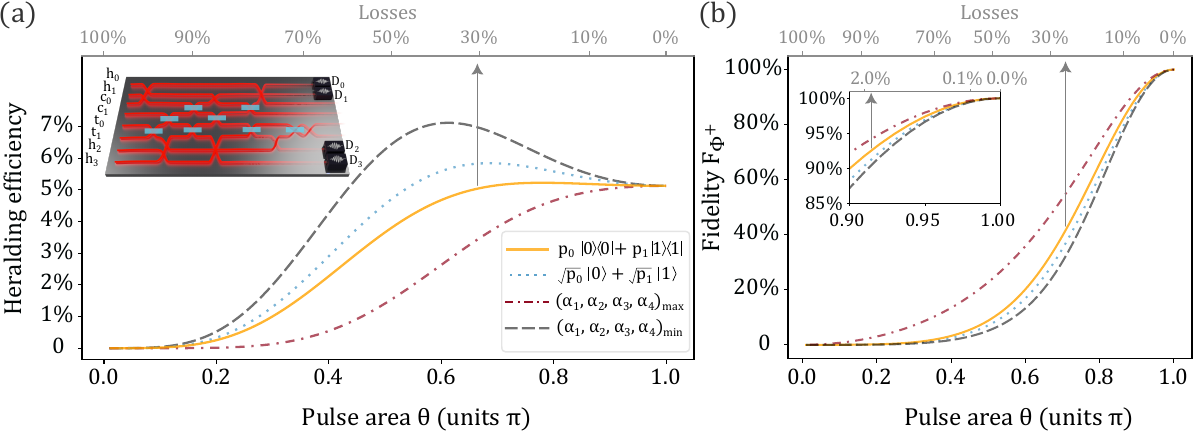}
\caption{(a) The heralding efficiency and (b) fidelity with respect to the Bell state $\ket{\Phi^+}$ as a function of pulse area $\theta$ of the driving field generating the photonic states input to the heralded CNOT gate (inset) where the target (t), control (c) paths encode logical qubits and certain combinations of detector clicks at the output of the heralding (h) paths signal a successful gate~\cite{Knill2001}. The different curves correspond to different scenarios: where physical input states share the same phase, $\alpha = 0$ (blue, dotted), where $\alpha$ is optimised to obtain maximum fidelity (red, dashed-dotted), and where $\alpha$ corresponds to a minimised gate fidelity (grey, dashed). Also shown are the heralding efficiency and fidelity as a function of losses (yellow, solid). \label{fig:6}  }
\end{figure*}

To illustrate this effect, we consider the case  of the heralded CNOT gate~\cite{Knill2001} in the path encoded implementation of Ref.~\cite{okamoto2018}, see inset Fig.~\ref{fig:6}(a). This gate requires four ancillary modes (h) to implement the non-linearity and herald the successful operation on the control (c) and target (t) qubit. The gate is heralded by the detection of exactly one photon at detectors D$_1$ and D$_3$ and zero photons at detectors D$_0$ and D$_2$.
If the qubits are prepared in the logical state $ \ket{\psi_\text{in}}=\frac{1}{\sqrt{2}}(\ket{0_l}_c + \ket{1_l}_c)\otimes \ket{0_l}_t$, the output logical state is a maximally entangled Bell state $\ket{\Phi^+} = \left(\ket{0_l}_c\ket{0_l}_t+\ket{1_l}_c\ket{1_l}_t\right)/\sqrt{2}$. The heralding probability, in the ideal case where none of the four photons gets lost, is $P_{(h|4)}= (11 - 6\sqrt{2})/49$, or approximately 5.1\%~\cite{okamoto2018}.

To model the effect of coherence with vacuum on the operation of this gate, we perform numerical simulations using the Perceval framework~\cite{Heurtel2022}. We first calculate the heralding efficiency $P_{(h)}$ considering incoherent losses, i.e. considering input states in the form of $\psi^\prime(\theta)=p_0 \ket{0}\bra{0} + p_1\ket{1}\bra{1}= \cos^2(\theta/2) \ket{0}\bra{0} + \sin^2(\theta/2)  \ket{1}\bra{1}$ (solid yellow curve in Fig.~\ref{fig:6}(a), and corresponding upper x-axis). We then consider the case of a quantum superposition of 0 and 1 photons. All four input single-photon states are initially prepared in the same state $\ket{\Psi(\theta,\alpha=0)}$ (according to Eq.~\ref{ref-eq1}) and we plot the corresponding heralding efficiency (dotted blue curve). 

We find that the maximal heralding efficiency can be increased from $5.1\%$ at $\theta=\pi$ to $5.8\%$ for a certain pulse area of $\theta<\pi$. This observation can be understood considering that coherence with vacuum leads to additional interference effects when the number of measurements is lower than the number of manipulated light pulses. As observed on the single counts and on the $|k|=1$ peaks in the HOM experiment, it can either reduce or increase the amplitude at certain outputs such that $P_{(h)}$ can be increased or decreased compared to the case of incoherent vacuum.  Actually, by numerically exploring $P_{(h)}$ when varying individually the four phases ($\alpha_1$,$\alpha_2$,$\alpha_3$,$\alpha_4$) of the four input states $\ket{\Psi(\theta, \alpha)}$, we find that $P_{(h)}$ can span anywhere from $2.5\%$ up to $7.1\%$ at $\theta=0.6\pi$ (dashed and dotted-dashed curves).

Conversely, Fig.~\ref{fig:6}(b) shows that the fidelity $F$ of the output state $\ket{\psi_\text{out}}$ with respect to the ideal Bell state is decreased when the four light states are in the input state $\ket{\Psi(\theta,\alpha=0)}$ (dotted blue) compared to the case of equivalent incoherent vacuum. Symmetrically, for the phase combination ($\alpha_1$,$\alpha_2$,$\alpha_3$,$\alpha_4$) that maximises (minimises) the heralding efficiency, we find that the fidelity of the heralded state is respectively decreased (increased) compared to incoherent losses. This behaviour can  be understood considering that errors arise only from additional vacuum components in the input (either coherent or incoherent). As a result, an enhanced heralding efficiency with respect to incoherent loss arising from quantum interference necessarily reduces the probability of obtaining two photons at the logical outputs. 

Quantitatively, the gate operates perfectly ($F=100\%$) with conditional probability $P_{(h|4)}$, i.e. a heralding signal was observed given that the input state contains exactly four photons, while $F=0\%$ if the input state contains fewer than four photons. In the case of either coherent or incoherent loss, this implies that the scheme will operate as expected only with probability $P_{(4)}={p_1}^4$ where $p_1=\sin^2(\theta/2)$. Prior to heralding, the fidelity of the output state thus depends only on the probability of having four photons at the input $P_{(4)}$ and the ideal gate heralding probability $P_{(h|4)}$. As such, the fidelity post-heralding is simply the probability $P_{(4|h)}$ of having had four photons given that a heralding signal was observed. Thus, using Bayes' theorem, we get
\begin{equation}
    F = \frac{P_{(4)} P_{(h|4)}}{P_{(h)}},
\end{equation}
where $P_{(h)}$ is the only term depending on the photon-number coherence and relative phases of the superposition with vacuum.

It is interesting to note that any suppression of the heralding efficiency $P_{(h)}$, stemming from coherence, acts as a filter to reduce the occurrence of erroneous output states that contain fewer than 2 photons, thereby enhancing the heralded Bell-state fidelity overall. Likewise, any enhancement of $P_{(h)}$ increases the number of erroneous output states, leading to a reduction in the fidelity.

\section{Conclusion  \protect}

We have shown that the photon-number coherence naturally present in the light wavepackets generated by quantum emitters leads to a large variety of quantum interference phenomena and entanglement. They impact {both the standard techniques employed in the development of deterministic quantum light sources and} information processing with photons. 
Starting with an experimental configuration as simple as the Hong-Ou-Mandel interferometer, we have shown that the long standardised protocol to deduce the photon indistinguishability {from quantum emitters needs to be  revised}

In the broader context of quantum information processing  { in the discrete variable framework, i.e. exploiting single-photon wavepackets}, our study  shows that the absence of a photon cannot simply be treated as a photon loss as has been done so far {in the discrete variable paradigm community}. {In particular, we experimentally demonstrate that the presence of photon-number coherence can be a resource to create entanglement in an unbalanced Mach-Zehnder interferometer, which touches on fundamental relationships in quantum physics~\cite{Killoran2016}.} As another example, we have found that coherence with vacuum can actually lead to reduced errors {for a photonic CNOT gate} compared to incoherent vacuum when leveraging control over the phase of the coherent superposition. {This hints that the amount and form of quantum coherence plays an important role in determining the performance of quantum information processing. Such a general relationship may be better elucidated from the perspective of resource theory~\cite{Baumgratz2014}.} 
We anticipate that the possibility to {deterministically} generate quantum superpositions of zero and one-photon in a fully controlled manner with quantum emitters opens up many possibilities for photon-based quantum information processing, providing additional degrees of freedom to leverage { single-rail qubit encoding in a revisited way of exploiting}. {It also may provide a critical bridge between continuous- and discrete-variable paradigms of quantum information processing. }

\begin{backmatter}
\bmsection{Funding}
This work was partially supported by the Agence Nationale de la Recherche (QuDICE project). We acknowledge funding from the Plan France 2030 through the projects ANR-22-PETQ-0011 and ANR-22-PETQ-0006, from the  H2020-FET OPEN project number 899544 - PHOQUSING, the French RENATECH network, the Paris Ile-de-France Region in the framework of DIM SIRTEQ. C. Anton-Solanas acknowledges the support from the Comunidad de Madrid fund “Atracci\'on de Talento, Mod. 1” 2020-T1/IND- 19785, Grant No. PID2020113445-GB-I00 funded by the Ministerio de Ciencia e Innovaci\'on (10.13039/501100011033), the grant ULTRA-BRIGHT from the Fundaci\'on Ramon-Areces, and the “Ambassade de France en Espagne".

\bmsection{Disclosures}
The authors declare no conflicts of interest.

\bmsection{Data availability} 
Correspondence and requests for data should be addressed to I.M.d.B.W. (imaillet@ic.ac.uk) or P.S. (pascale.senellart-mardon@c2n.upsaclay.fr). 

\bmsection{Supplemental document}
See Supplement 1 for supporting content.

\end{backmatter}


\bibliography{references}

\end{document}